\documentclass[preprintnumbers,amsmath,amssymb,prc,showpacs]{revtex4} 
\usepackage[english]{babel}
\usepackage[dvips]{graphics,graphicx}
\usepackage{epsfig}
\usepackage{amssymb,amsmath,amsthm,amsfonts,amssymb}

\begin{document}

\title{Energy Extraction from the Einstein-Born-Infeld Black Hole}
\author{Nora Bret\'on }
\affiliation{
$^{}$ Physics Department, Centro de Investigaci\'on y de Estudios Avanzados 
del I. P. N. (Cinvestav), Mexico City.}

\begin{abstract}
The energy extraction from a Einstein-Born-Infeld (EBI) black hole is addressed determining the extension of the ergosphere as well as the extractable energy using the irreducible mass concept. These results are  compared with the Reissner-Nordstrom (RN) ones; RN is the linear electromagnetic counterpart of the BI black hole. It turns out that for a fixed charge $Q$, more energy can be extracted from the RN black hole than from the EBI one. The extreme case is investigated as well, presenting remarkable features, for instance that more energy can be extracted from extreme EBI black holes  than from extreme RN, however, extreme EBI black holes lack of a linear electromagnetic black hole limit.
\end{abstract}


\maketitle


\section{Introduction}

Energy extraction from black holes has interest in its own as fundamentals of black holes but also as an astrophysical
engine sourcing  large energetic jets or gamma ray bursts. Several black hole extraction processes are known: Penrose process,
magnetohydrodynamics Penrose process, superradiance, among others.

By means of the Penrose process energy can be extracted from a rotating black hole through particle fission: initially a particle enters the ergosphere and then splits into two particles, one of them, on a negative energy orbit, falls into the spinning black hole, the other escapes to infinity with more energy than the original particle, while the black hole losses some of its angular momentum. The process relies on the existence of a region, outside the horizon, where negative energy orbits are allowed but such that  particles still can escape from black hole. This region is called the ergosphere. In Kerr black hole the ergosphere extends from the horizon to the static limit. Besides the rotating black holes, energy can be extracted from charged black holes \cite{Christodou1971}, as a result of the black hole charge interaction with the physical parameters of the interacting particles or fields. Denardo and Ruffini (1973) \cite{Denardo1973} defined this region as the \textit{effective ergosphere}. The extension of the effective ergosphere depends on the sign and magnitude of the test charge. The static and spherically symmetric solution of the coupled Einstein-Maxwell equations is the  Reissner-Nordstrom (RN) black hole, being the so called extreme RN black hole the one with its mass equating its charge, $M=Q$.
It turns out that up to the 50 \% of the total charge of an extreme RN  black hole can be extracted.
Being charged black holes so a promising energy extraction source, it is important to determine how much energy can be extracted from other than RN black holes.

The existence of astrophysical charged black holes has been questioned, however black holes with fields of the order of the critical field to  polarize vacuum can be important in finding an explanation for ultra high energy cosmic rays \cite{Vitagliano2002} or the modeling of gamma-ray bursts \cite{Preparata1998}.  
A very high energy electromagnetic field fails to obey Maxwell's electrodynamics and to give a reliable description of such scenarios quantum electrodynamics (QED) should be employed.
In this direction we are addressing the Einstein-Born-Infeld (EBI) black hole. Born-Infeld (BI) electrodynamics \cite{BI1934} outstands among nonlinear electromagnetic (NLEM) theories  for several reasons, namely, BI  can be considered as an effective theory that takes into account quantum electromagnetic effects, for instance vacuum polarization or light by light dispersion, to the tree-level approximation in QED \cite{Schwinger1951}. Another reason is that BI lagrangian arises in the low-energy limit in string theory and BI solutions can be interpreted as D-branes \cite{Gibbons1998}. Moreover, on a D3-brane, open strings attached to the brane may couple to a $U(1)$ field at the end of the string while the string tension is related to the BI parameter \cite{Tseytlin1985}.

In this paper the energy extraction from a EBI black hole  is analyzed determining the extension of its ergosphere as well as the extractable energy using the irreducible mass concept; the NLEM behavior is compared with the Einstein-Maxwell one.  It turned out that the energy that can be extracted from the EBI black hole is alwyas less that the one extractable from RN black hole, having the former as its upper limit the latter. The extreme black hole is investigated as well, presenting remarkable features; in principle more energy can be extracted from extreme EBI black hole than from extreme RN, however the lacking of a linear limit of the extreme BI black hole may be an inconvenient. Some comments are included about another energy extraction process, namely, superradiance from the EBI black hole.


\section{The  Einstein-Born-Infeld  Black Hole}

Born and Infeld (1934)  \cite{BI1934}   proposed a  Lagrangian that makes finite the electromagnetic field
at the charge position, introducing the parameter $b$  as the maximum allowed electromagnetic field. The BI Lagrangian is given by

\begin{equation}
{\cal L}_{\rm BI}= 4 b^2 \left({ -1 + \sqrt{1+ \frac{F}{2 b^2}+ \frac{G^2}{16b^4}}}\right),
\end{equation}
where the electromagnetic invariants $F$ and $G$ are
$F=F_{\mu \nu}F^{\mu \nu}= 2(B^2-E^2), \quad   G={\tilde F}_{\mu \nu}F^{\mu \nu}= \vec{E} \cdot {\vec{B}}; {\tilde F}_{\mu \nu}$ is the dual field-strengh electromagnetic tensor,  $ {\tilde F}^{\mu \nu}= \epsilon^{\mu \nu \alpha \beta} F_{\alpha \beta} ;$
$b$ is the maximum allowed field, whose value was  calculated as  $b=10^{20}$Volt/m by considering that mass has its origin in the electromagnetic field, adopting the unitarian standpoint of view, which postulates
 the existence of only one physical entity  \cite{BI1934}.
The linear limit is the  Maxwell Lagrangian,  given by $ {\cal L}_{\rm Max}=F.$

Very soon after the BI proposal came up it was realized that
Euler-Heisenberg and Schwinger \cite{Schwinger1951} Lagrangians in QED are related to BI Lagrangian, more precisely, QED Lagrangian at the tree-level expansion  can be derived in the weak-field limit from the BI Lagrangian i.e. BI Lagrangian may be considered an effective Lagrangian of QED vacuum polarization.

The Einstein-Born-Infeld (EBI) action is given by 

\begin{equation} 
S=  \frac{1}{16 \pi} \int{d^4x \sqrt{g} \left( R - {\cal L}_{\rm BI}\right)}. 
\end{equation}

The  static and axisymmetric solution of EBI action is given by the metric function and the electromagnetic potential as follows  \cite{GSP1984},
  
\begin{eqnarray}
ds^2&=&\psi(r)dt^2-\psi^{-1}(r)dr^2-r^2d\Omega^2, 
\label{SSSm}\\ 
\psi=\psi^{\rm BI} &=&1- \frac{2M}{r}+ \frac{2}{3}r^2 b^2 \left({1-\sqrt{1+ \frac{Q^2}{ \beta^2 r^4}}}\right) + \frac{4Q^2}{3r} I(r),
\label{BIsol}\\
A_{t}(r)&=& Q I(r),
\end{eqnarray}
where
\begin{equation}
I(r)=\int_{r}^{\infty}{\frac{dz}{\sqrt{z^4+Q^2/b^2}}}= \frac{1}{2} \sqrt{\frac{b}{Q}} \emph{F} \left[{\arccos\left({\frac{r^2-Q/b}{r^2+Q/b}}\right)},\frac{1}{\sqrt{2}}\right],
\label{BIint}
\end{equation}
where $ \emph{F}$ is the Legendre elliptic function of the first kind. The integral can also be given in terms of the Hipergeometric function when the argument is less than one. In the previous expression  due to the duality of BI theory \cite{Gibbons1995} the charge may be electric, magnetic, or both, $Q^2=Q_m^2+Q_e^2$.  The electromagnetic field at the origin  $E(r=0)$ is finite, although there are curvature singularities at $r=0$.
The solution may present  one, two or none  horizons depending on the balance between the parameters of mass, charge and BI parameter, $M,Q,b$.
 
The metric function in Eq. (\ref{BIsol}) can be written as \cite{Breton2002},

\begin{equation}
\psi^{\rm BI} = 1+ \frac{2}{3} b^2 \left({r^2-\sqrt{r^4+ \frac{Q^2}{ \beta^2 }}}\right) - \frac{2}{r} \left({M - \frac{2Q^2}{3} \int_{r}^{\infty} \frac{dz}{ \sqrt{z^4+q^2/ b^2}} }\right).
\end{equation}

From the previous expression notice that if the last term is zero at the origin, $r=0$, a regular metric is obtained. However there persists the curvature singularity at the origin, for this reason these BHs have been called marginal \cite{Mann2012}, \cite{Fernando2013}, \cite{Gibbons2003}. This case occurs for $M=M_0$ where 

\begin{equation}
M_0=\frac{2Q^2}{3} \int_{0}^{\infty} \frac{dz}{ \sqrt{z^4+q^2/ b^2}}=\frac{2}{3}Q^{3/2}b^{1/2} K \left[\frac{1}{\sqrt{2}}\right],
\end{equation}
where $ \emph{K}$ is the complete Legendre elliptic function of the first kind. 
For values of the gravitational mass such that $M>M_0$ the dominat behavior is Schwarzschild-like, this meaning that only one horizon is present \cite{Mann2012}. From these black holes no energy can be extracted, since every particle that crosses the horizon has no possibility to get off the black hole, i.e. in those cases there is no ergosphere. 
 
In the cases that $M<M_0$,  a RN-like behavior dominates existing two horizons.
In the limit $ b \to \infty$ the Reissner-Nordstrom solution is recovered,

\begin{equation}
\psi^{\rm RN}(r)= 1 - \frac{2M}{r}+ \frac{Q^2}{r^2}, \quad
A_{t}= \frac{Q}{r}.
\end{equation}
  
The RN outer and inner horizons, $r_{\pm}$ are given by the solutions of  $\psi^{\rm RN}(r)=0$,

\begin{equation}
r^2 - {2M}{r}+ Q^2=0, \quad r_{\pm}= M \pm \sqrt{M^2-Q^2}.
\end{equation}

The extreme RN black hole is when $M^2=Q^2$, in this case the two horizons coalesce into one.
The extreme RN black hole is of great importance in string theory as a BPS state, that is a classical state that preserves quantum symmetries.

\begin{figure}
\centering
\includegraphics[width=14cm,height=6cm]{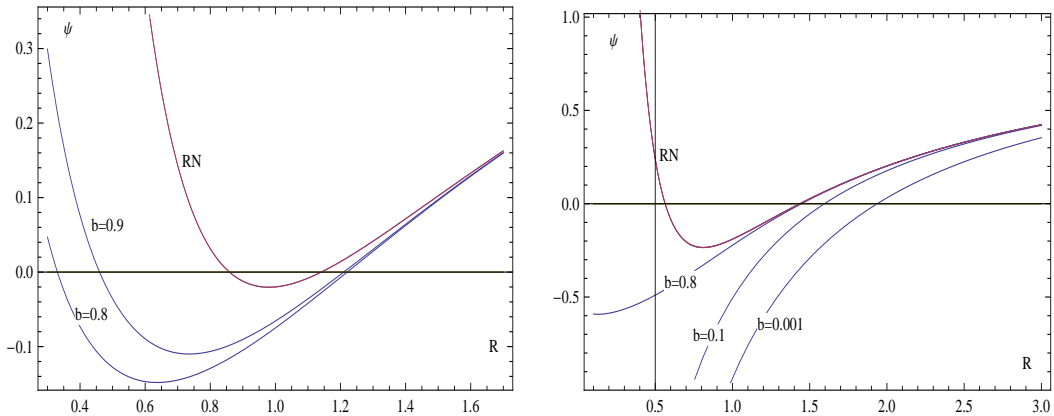}
\caption{\label{fig1} 
The  EBI and RN metric functions, as function of $R=r/M$, are compared for fixed values of  charge, $Q=0.99$ and mass$M=1$, varying the BI parameter $b$, whose values are on the corresponding curve. As $b$ gets larger the EBI metric function approaches the RN one.}
\end{figure}

In Fig. \ref{fig1} are compared  the  EBI and RN metric functions for fixed values of charge, $Q=0.99$ and mass$M=1$, varying the BI parameter $b$, whose values are on the corresponding curve. As $b$ gets larger the EBI metric function approaches the RN one. Note that the EBI horizon is always larger than the RN one.  Due to presence of  the nonlinear electromagnetic field there is an  effect of charge shielding that results in a black hole with a larger horizon than the  RN horizon. Remind that the charged black hole is a more compact object than Schwarzschild. From hereafter ``black hole" may be abbreviated by BH.

\subsection{The EBI black hole ergoregion}

For a charged black hole the ergoregion is a  region where  the 4-momentum of  a test particle is spacelike \cite{Christodou1971}.
To determine the extent of the EBI-BH ergoregion we analize the charged test particle trajectories. 
In a static axisymmetric  spacetime of the form (\ref{SSSm}) the energy $E$ and angular momentum $L_z$ of a test particle of mass $\mu$ and charge $q$ are conserved quantities,

\begin{equation}
p_{t}=E=- \mu \psi \dot{t}-q A_{t}, \quad p_{\varphi}=L_{z}= \mu g_{\varphi \varphi} \dot{\varphi}+ q A_{\varphi},
\end{equation}
 
that allow the integration of the motion equations,

\begin{equation}
\psi \dot{t}^2 - \psi^{-1} \dot{r}^2 - r^2 \dot{\theta}^2- \sin^2{\theta} \dot{\varphi}^2=1.
\end{equation}

Restricting the orbits to the equatorial plane $\theta= \pi/2$ and at the turning point $\dot{r}=0$, substituting the conserved quantities, it is obtained

\begin{equation}
\left({\frac{E}{\mu}-\frac{q}{\mu}A_{t}}\right)^2=\psi \left({1+ \frac{p_{\varphi}^2}{\mu^2 r^2}}\right).
\end{equation}
 
The previous equation is quadratic on the energy of the test particle $E$,  

\begin{equation}
 \left(\frac{E}{\mu} \right)^2 -2 \left(\frac{q}{\mu} \right)A_{t} \left(\frac{E}{\mu} \right)+ \left(\frac{q}{\mu} \right)^2A_{t}^2-  \psi \left(1+ \frac{p_{\varphi}^2}{\mu^2 r^2}\right)=0, 
\label{Ergoeq}
\end{equation}

the zeroes of such a polynomial being,

\begin{equation}
E_{\pm}= qA_{t} \pm \mu \left(\psi \left[1+ \frac{p_{\varphi}^2}{\mu^2 r^2}\right]\right)^{1/2}.
\label{Ergoenergy}
\end{equation}

\begin{figure}
\centering
\includegraphics[width=8cm,height=6cm]{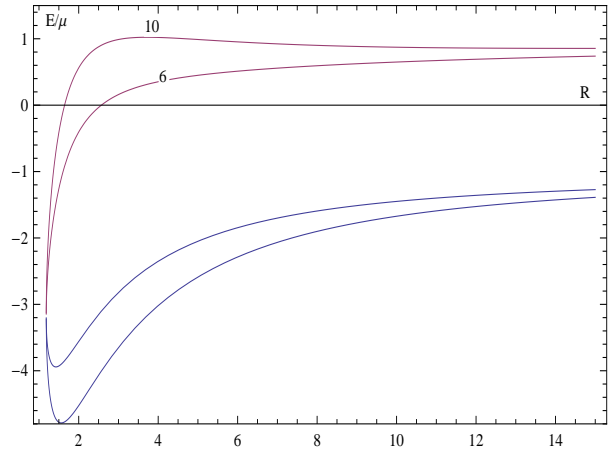}
\caption{\label{fig2}
It is shown $E/ \mu$ from Eq. (\ref{Ergoenergy}) for a charged test particle in the BI spacetime. 
The regions with negative energy may be thought as an effective ergosphere from which energy can be extracted.
Upper curve corresponds to  $p_{\varphi}=10$ while the lower one is for  $p_{\varphi}=6$. The rest of parameters fixed as $M=1, Q=1,b=0.8, q/ \mu=-4$ and $R=r/M$.}
\end{figure}

 $E_{+}$ corresponds to test charge 4-momentum pointing towards future. If $A_t >0$,
for $q<0$ negative energy $E_{+}$ states exist and then energy can be extracted. This is illustrated in Fig. \ref{fig2}.
 From Eq. (\ref{Ergoenergy}) for $E$,  the orbit with minimum energy is very close to the horizon, denoted by $r_{+}$, where $\psi(r_{+})=0$ and the square root is zero, 

\begin{equation}
E_{+}=qA_t(r_{+}).
\end{equation}

Note that no extraction is possible in the limit $p_{\varphi} \to \infty$, and that the most negative energy  $E_{+}$, outside of the horizon,  is reached when $p_{\varphi}=0$,

\begin{equation}
E_{\pm}= qA_{t} \pm \mu \left(\psi \right)^{1/2}.
\label{EnergyEq}
\end{equation}

The largest radius $r_{\rm es}$ that is determined from  $E_{+}=0$ in Eq. (\ref{EnergyEq}), 
defines the outer limit of the ergosphere,  the inner radius being the horizon.

For the RN case, the ergosphere includes the RN horizon up to the largest root of $E_{+}=r^2-2Mr+Q^2[1-q^2/ \mu^2]=0$, 

\begin{equation}
r_{+}^{\rm RN}=M+ \sqrt{M^2-Q^2} \le r \le M+ \sqrt{M^2-Q^2\left({1- \frac{q^2}{\mu^2}}\right)} =r_{\rm es}^{\rm RN}.
\end{equation}
 
In the EBI case the equation for the horizon $r_{+}^{\rm BI}$ is more complicated while  $r_{\rm es}^{\rm BI}$ is given by the largest root of

\begin{equation} 
r-2M+\frac{2}{3} b^2r(r^2-\sqrt{r^4+Q^2/b^2})+ Q^2I(r) \left({\frac{4}{3}-r{\frac{q^2}{\mu^2}}I(r)}\right)=0,
\label{r_esBI}
\end{equation}
 with $I(r)$ from Eq. (\ref{BIint}).
\begin{figure}
\centering
\includegraphics[width=12cm,height=6cm]{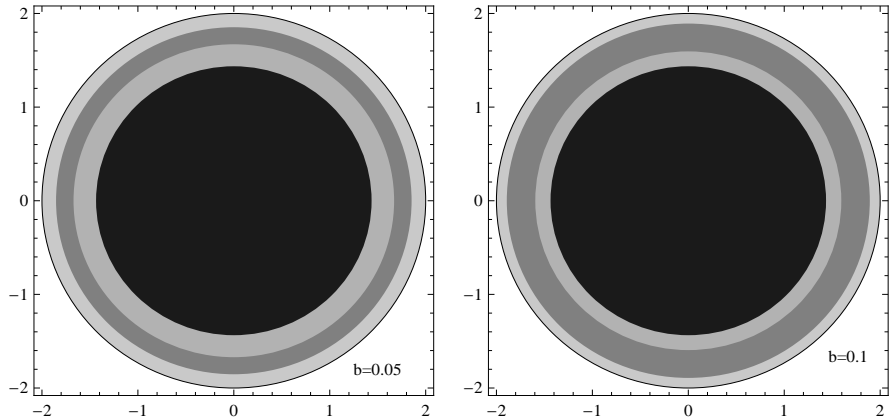}
\caption{\label{fig3}
Dark grey ring (spherical shell) is the EBI ergoregion that is always smaller than the one of RN (light grey); 
as $b$ grows, the BI ergoregion approaches the RN one that is its upper limit. The black disk represents the RN horizon.
In the plot $M=1,Q=1,b=0.05$ (left); $b=0.1$ (right), and the charge test parameters are $\mu=1, q=-1.$}
\end{figure}

Taking  the linear limit, $b \to \infty$ in Eq. (\ref{r_esBI}), expanding the square root and considering that in that limit $I(r)=1/r$, Eq. (\ref{r_esBI}) becomes

\begin{equation} 
r^2-2Mr+Q^2 \left({1-\frac{q^2}{\mu^2}}\right)=0,
\end{equation}
whose solution is $r_{\rm es}^{\rm RN}$,  showing that as $b$ increases the EBI ergosphere approaches the RN one, being the latter its upper limit. This is shown in Fig. \ref{fig3}  where the shells of the EBI and RN ergospheres are compared for two values of the BI parameter $b$.

\subsection{The irreducible mass and the extractable energy}

Extractable energy from a black hole can be formulated using the irreducible mass concept.   It is defined as the quantity that cannot decrease through a reversible process. In such a process gravitational radiation should be negligible.  Moreover,  the irreducible mass concept is deeply connected to the first and second  laws of black hole mechanics \cite{Rasheed97}.

It has been shown that for a spherically symmetric black hole within nonlinear electromagnetic fields \cite{Pereira2014} the irreducible mass is given in terms of the horizon radius $r_{+}$  as

\begin{equation}
M_{ir}=\frac{r_{+}}{2}.
\label{Mir}
\end{equation}

For the RN black hole the  $M_{\rm ir}$ is given by

\begin{equation}
 M_{\rm ir}^{\rm RN}=\frac{1}{2}(M+ \sqrt{M^2-Q^2}),
\end{equation}
while $M_{\rm ir}^{\rm BI}= \frac{1}{2}r_{+}^{\rm BI}$ with  $r_{+}^{\rm BI}$ being the largest root of  $\psi^{\rm BI} (r)=0$, 

\begin{equation}
\frac{2}{3} b^2r(r^2-\sqrt{r^4+Q^2/b^2})+ \frac{4}{3}Q^2I(r)+ r-2M=0.
\end{equation} 

\begin{figure}
\centering
\includegraphics[width=14cm,height=6cm]{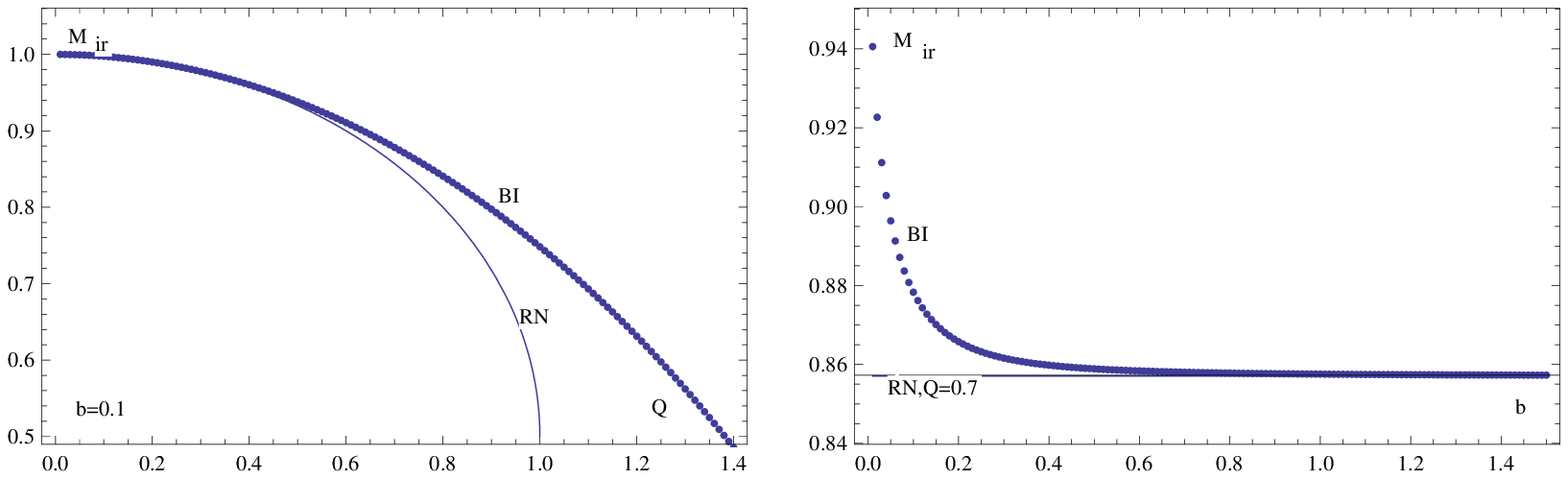}
\caption{\label{fig4}}
The irreducible mass $M_{\rm ir}=r_{+}/2$ of both the RN and EBI BHs is shown for a fixed mass $M=1$, to the left with fixed  BI parameter $b=0.1$, while varying electromagnetic charge $Q$. The admissible charge range for RN is 
$0 \le Q \le 1$ corresponding the maximum charge to the extreme RN BH, for which the minimum $M_{\rm ir}=0.5Q$ is reached.
 To the right EBI and RN irreducible mass is compared for a fixed mass $M=1$ and charge $Q=0.7$ and varying the BI parameter $b$.
In both cases it is fulfilled that RN-$M_{\rm ir}$ is smaller or equal  than  BI
-$M_{\rm ir}$.
\end{figure}

In Fig. \ref{fig4} are displayed
the irreducible mass of both the RN and EBI BHs  for a fixed mass $M$,
alternatively  fixing the  BI parameter $b=0.1$, while varying electromagnetic charge $Q$ and then  fixing the charge and varying $b$.
In both cases it is fulfilled that RN-$M_{\rm ir}$ is smaller than  BI-$M_{\rm ir}$. Then Eq. (\ref{Mir}) implies that the RN horizon is  smaller or equal than the BI one,
$r_{+}^{RN}<r_{+}^{\rm BI}$. Besides, since  $M_{\rm ir}^{\rm RN}<M_{\rm ir}^{\rm BI}$ then more energy can be extracted from the RN BH than from the BI one.

 In \cite{Pereira2014} it was shown that the total mass-energy of  a spherically symmetric black hole within nonlinear electromagnetic theory can be decomposed in terms of the irreducible mass. Such decomposition is given by

\begin{equation}
M=M_{\rm ir}+ 4 \pi \int_{r_{+}}^{\infty}{x^2 T_{0}^{0}(x) dx}.
\end{equation}

The extractable energy $E_{\rm extr}$ turns out to be the ADM mass minus the irreducible mass \cite{Pereira2015}

\begin{equation}
E_{\rm extr}= M-M_{\rm ir} =4 \pi \int_{r_{+}}^{\infty}{x^2 T_{0}^{0}(x) dx}.
\end{equation}

In terms of the irreducible mass, the minimum of $M_{\rm ir}$ corresponds to the maximum of the extractable energy. 
That is, from Schwarzschild BHs no energy can be extracted, or in other words, for charged static BHs, the extractable energy comes out from the electromagnetic field. 

 For the EBI BH with an electromagnetic energy density of $T_{0}^{0}(r)= 2b^2 [ \sqrt{1+ Q^2/(b^2r^4)}-1]$, the extractable energy amounts to
 
\begin{equation}
E_{\rm extr}^{\rm BI}=\frac{b^2}{3}r_{+} [r_{+}^2- \sqrt{r_{+}^4+Q^2/b^2}]+ \frac{2}{3} Q^2  I(r_+).
\end{equation}
 
In the limit $b \to \infty$  the RN case is recovered,

\begin{equation}
E_{\rm extr}^{\rm RN}=\frac{Q^2}{2r_{+}}= \frac{Q^2}{2(M+ \sqrt{M^2-Q^2})}.
\label{extrRN} 
\end{equation}

\begin{figure}
\centering
\includegraphics[width=8cm,height=6cm]{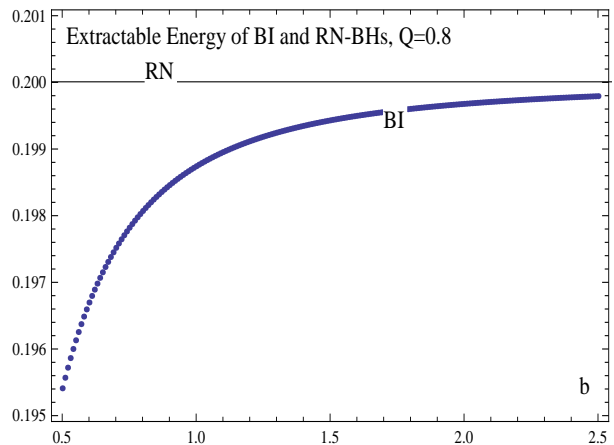}
\caption{\label{fig5} The behavior of the maximum extractable energy of the EBI-BH is shown,  as the BI parameter is varied, for a fixed mass and charge $M=1, Q=0.8$. It is compared to the maximum extractable energy from the RN-BH, $E_{\rm max}^{\rm RN}=0.2$ (from Eq. (\ref{extrRN})). As $b$ grows,
$E_{\rm extr}$ approaches the maximum $E_{\rm extr}$ of the RN-BH.
 }
\end{figure}

In Fig. \ref{fig5} the behavior of the extractable energy of the EBI-BH, as the BI parameter is varied, is shown for a fixed charge $Q$. It is compared to the maximum extractable energy from the RN-BH from Eq. (\ref{extrRN})  (for a fixed value of $Q,M$). This results hold for the non extreme BHs. The extreme cases deserve a particular analysis that will be addressed in the next subsection.

\subsection{The extreme Born-Infeld black hole and its extractable energy}

The extreme black hole is the case when the two horizons coalesce into one. The conditions that define it are that at the degenerate horizon both the metric function and its first derivative are zero. Imposing both conditions to the BI metric function $\psi (r)$ in Eq. (\ref{BIsol}), leads to the value of the degenerate horizon as

\begin{equation}
r_{+}^{\rm eBI}= \frac{\sqrt{4b^2Q^2-1}}{2b},
\end{equation}
the superindex eBI denotes the extreme EBI-BH; in principle it does not depend on the mass, but recall that $M<M_0$ in order to have two horizons, in this case coalesced into one. It turns out  that  $r_{+}^{\rm eBI}$ is smaller than  the extreme RN horizon,  $r_{+}^{\rm eRN}=Q=M$,

\begin{equation}
r_{+}^{\rm eBI}= \frac{\sqrt{4b^2Q^2-1}}{2b}=  Q \sqrt{1-\frac{1}{4Q^2b^2}} < Q =r_{+}^{\rm eRN}.
\end{equation}
  
If the last inequality is taken to the letter it would imply  that  the BI irreducible mass is smaller than the RN one, both in the extreme case, $M_{\rm ir}^{\rm BI} < M_{\rm ir}^{\rm Rn}$, and it  would lead to the conclusion that more energy can be extracted from the extreme EBI-BH than from the extreme RN-BH, which is inconsistent with  the previous results, and in contradiction with the fact that any horizon $r_{+}^{\rm BI} >r_{+}^{\rm RN}$ and then for continuity, $r_{+}^{\rm eBI} \ge r_{+}^{\rm eRN}$. 

\begin{figure}
\centering
\includegraphics[width=8cm,height=6cm]{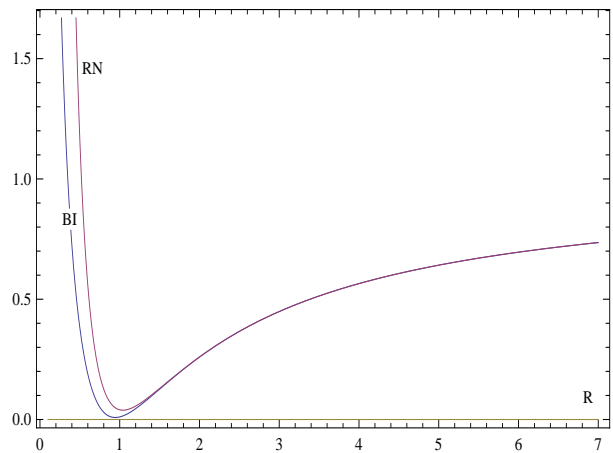}
\caption{\label{fig6} 
It is illustrated the metric functions of  an extreme EBI-BH showing the two coalesced horizons  and its RN linear limit, for a fixed value of the charge, $Q=1.02$. The mass has been fixed as $M=1$, then the linear counterpart, the RN solution is a naked singularity, i. e. the zeroes  of $\psi^{\rm RN}=0$ are  complex numbers.}
\end{figure}

\begin{figure}
\centering
\includegraphics[width=8cm,height=6cm]{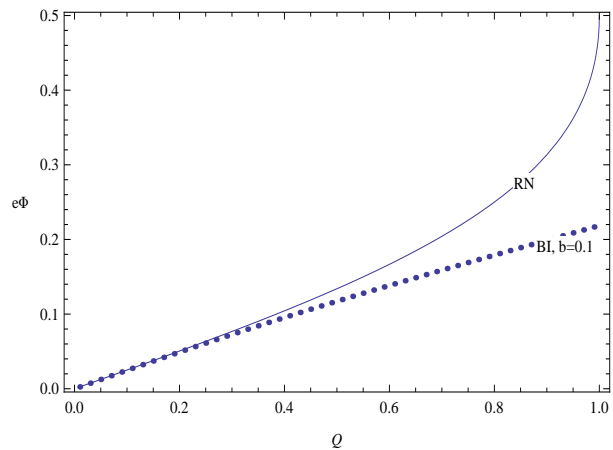}
\caption{\label{fig7} 
The frequency ranges $\omega$ of a scattering charged field impinging the EBI and RN BHs such that superradiance occur are shown; BI parameter $b$ and test charge are fixed as $b=0.1$ and $e=q=0.5$,  varying the BH charge $Q$. The area below the curves fulfils the superradiance condition, 
$\omega < e \Phi (r+)$. The BI range is smaller than the RN one.}
\end{figure}

Extreme EBI-BHs indeed exist;  by fixing $M$, the parameters $b$ and $Q$ can be balanced in order to have  extreme EBI-BHs; however, for a given choice of the mass, the charge should be greater than that fixed value, $Q>M$, and in that case the RN counterpart is not  a black hole anymore because its horizons that are the roots of $\psi^{\rm RN}=0$ are complex numbers, being the solution  a naked singularity. Therefore the extreme EBI-BHs lack of  BH linear limit. Since a  linear limit BH should exist as the object to which EBI-BH decays  in a discharging process, such extreme EBI-BH should be rejected.
 In Fig. \ref{fig6} an extreme EBI-BH and its corresponding linear limit, a naked singularity, are shown.

\section{Comments on Superradiance from EBI Black Hole}

Another extracting black hole energy process is superradiance scattering:  rotational energy may be extracted from a rotating BH by scattering waves upon it. Such waves are amplified upon scattering,  being then a way of extracting energy from the hole. The amplified wave should be of certain frequencies $\omega$, obeying the condition $\omega < m \Omega_{+}$ where $ \Omega_{+}$ is the horizon angular velocity and $m$ is the azimuthal quantum number of the wave mode. The analogue happens to charged BHs:  test charged fields interacting with RN black holes exhibit fast growing superradiant instabilities \cite{Bekenstein1973}. Charged waves with frequency $\omega$ impinging on a RN-BH are amplified provided  $\omega < q \Phi_{+}$, 
where  $\Phi_{+}$ is the electromagnetic potential evaluated at the horizon and $q$ is the charge of the test field. Stability of the process has been the subject of recent numerical studies \cite{JCDegollado2014}.

 For the RN black hole the superradiant condition amounts to

\begin{equation}
\omega < \frac{qQ/M}{1+ \sqrt{1-(Q/M)^2}},
\end{equation}
while for EBI-BH the superradiance condition becomes

\begin{equation}
\omega <  q \frac{\sqrt{Qb}}{2} \emph{F} \left[{\arccos \left({\frac{r_{+}^2-Q/b}{r_{+}^2+Q/b}}\right), \frac{1}{\sqrt{2}}}\right].
\end{equation}
 
The expression is
cumbersome and  $r_{+}^{\rm BI}$ is known only numerically, therefore the best way of getting some insight on the relation is by plotting the allowed range for the superradiance frequencies. It is shown  as the area below the curve in Fig. \ref{fig7}.  RN diverges at $Q=M$ that corresponds to the extreme RN-BH. In the plot for fixed $b$ and varying $Q$  it can be seen that the EBI superradiant frequency range is smaller that the one for RN, $\omega < q \Phi_{+}^{\rm BI}  < q \Phi_{+}^{\rm RN}$, giving then a stronger restriction on the field frequencies able to produce superradiance. 

\section{Final Remarks}

The problem of energy extraction from the nonlinear electromagnetic Born-Infeld (EBI) black hole has been addressed and results are compared to its linear electromagnetic counterpart, the Reissner-Nordstrom (RN) black hole. The EBI black hole is characterized by three parameters,  mass $M$, electric and magnetic charge $Q$ and the BI parameter $b$. The extent of the region from which energy can be extracted, the ergoregion, has been determined, as well as the irreducible mass, defined as such energy that cannot decrease through any irreversible process. 
The EBI ergoregion is smaller than the RN one; as $b$ increases the EBI ergoregion approaches the RN one, as its upper limit. 
In agreement with the ergoregion extent, the equivalent result in terms of the irreducible mass is that the BI irreducible mass is greater than the RN one; therefore less energy can be extracted from EBI-BH than from the RN one.
Irreducible mass of the RN-BH has a lower limit that corresponds to the extreme BH ($M=Q$). Regarding the extreme EBI-BH it was proved that
it exists and is not unique; fixing the mass, then the charge $Q$ and the BI parameter $b$ can be arranged such that the two horizons coalesce into one for several values of $(Q,b)$.  For a given  mass  $M$, there are many  extreme EBI-BHs but all of them for  $Q>M$; consequently,  in such cases the RN counterpart is a naked singularity, because RN horizon turns out to be complex, since the square root is imaginary, $r_{+}^{\rm RN}=M+ \sqrt{M^2-Q^2}$. Hence extreme EBI-BHs do not have a BH  linear counterpart, in which case its mere existence may be questionable on physical grounds.

Finally, regarding energy extraction by means of the superradiance process, it was determined that the frequency range of the impinging wave upon the EBI-BH for the occurrence of superradiance is smaller than the corresponding to RN-BH.  

In summary, it has been  shown that the introduction of a nonlinear electromagnetic field of the BI kind reduces the black hole ability to render energy through extraction processes based on its charge, as compared with RN black hole. If we interpret nonlinear electromagnetic field as equivalent to  embedding the black hole  into a material medium, that is the most likely astrophysical scenario, we cannot be so optimistic as to consider energy extraction from electromagnetic charge, in terms of efficiency, to play an important role among extracting energy processes. These results hold for QED to the tree-level approximation, being the Euler-Heisenberg Lagrangian a limit  of the Born-Infeld Lagrangian for low energy electromagnetic fields \cite{Schwinger1951}.



\end{document}